\documentclass{aa}

\usepackage{txfonts}
\usepackage{graphicx}
\usepackage{natbib}
\usepackage{amssymb}

\bibpunct{(}{)}{;}{a}{}{,} % to follow the A&A style

\begin{document}
\title{Can $\Lambda$ be determined from nearby Type Ia Supernovae?}
\subtitle {}
\author   {P. Erni \inst{1} and G.A. Tammann \inst{2}}
\institute{Argelander-Institut f\"ur Astronomie\thanks{Founded by merging of the Institut 
           f\"ur Astrophysik und Extraterrestrische Forschung, the Sternwarte, and the 
	   Radioastronomisches Institut der Universit\"at Bonn.}, Universit\"at Bonn,
           Auf dem H\"ugel 71, 53121 Bonn, Germany	
     \and  Astronomisches Institut der Universit\"{a}t Basel, Venusstrasse 7, CH-4102 Binningen, Switzerland   
     }
     
\date{ }

\titlerunning{Can $\Lambda$ be determined from nearby Type Ia Supernovae?}
\authorrunning{P. Erni and G.A. Tammann}

\abstract{
Type Ia Supernovae (SNe~Ia) are the best standard candles known today. At high
redshift ($z\sim1$) SNe~Ia are used to determine the Cosmological Constant
$\Lambda$ with great success. However the most serious concern is raised by the
possible luminosity evolution of the SNe~Ia explosion itself, i.e., that their
intrinsic luminosity might vary with the look-back time. It is unknown to which
extent high-redshift SNe~Ia can directly be compared to near SNe~Ia in order to
determine $\Lambda$. A possibility to circumvent this problem is to restrict the
analysis to nearby SNe~Ia situated preferably in E/S0 galaxies. Since the signal 
will be much smaller, we have to consider an substantial sample. As there are not 
enough data yet available, we conducted our analysis based on 200 synthetic SNe~Ia 
with a luminosity scatter $\sigma_{\mathrm{m}}=0.^{\mathrm{m}}12$ (derived from 
observations) assuming a homogeneous space distribution and a limiting distance
of $z\leq0.16$. We show that this kind of data, which we expect from future observations, 
will allow us to distinguish between a matter dominated or $\Lambda$-dominated 
($\Omega_{\mathrm{M}}=0.3$, $\Omega_{\mathrm{\Lambda}}=0.7$) universe with a 
significance of up to 2-3\,$\sigma$.

}

\maketitle
\section{Introduction}
The original hope of using HST to determine the Cosmological
Constant $\Lambda $ and the matter density parameter $\Omega _{\mathrm{M}}$ from
high-redshift standard candles, as exemplified by SNe~Ia \citep{tammann1979}, was
substantiated by two brilliant experiments of independent groups using SNe~Ia with
redshifts of up to $z\sim0.8$ \citep{riess1998,perlmutter1999}.
The significant conclusion from these was that $\Lambda $ is positive for any
value of $\Omega _{\mathrm{M}}$. Various effects have been discussed which could
alter this result, e.g., ad hoc postulated gray absorption or strong and weak
gravitational (anti-) lensing \citep{valageas2000}. Yet the most severe question
is raised by the possibility of luminosity evolution of SNe~Ia, i.e., that
their intrinsic luminosity at maximum vary with the look-back time.

The determination of $\Lambda$ from nearby SNe~Ia \citep[see also][]{germany2004}, 
say at $z\leq0.16$, is therefore an intriguing possibility. 
In the Hubble diagram the $\Omega_{\mathrm{M}}=1$ model 
is separated from an $\Omega_{\mathrm{M}}=0.3$, $\Omega_{\Lambda}=0.7$ model by 
$0.^{\mathrm{m}}53$ at $z=0.8$, but only by $0.^{\mathrm{m}}15$ at $z=0.16$. In the
latter case the advantage is that the look-back time is reduced by roughly a factor
of 3 to 2.3 Gyr. The decisive question for the present experiment is therefore 
whether luminosity evolution can be excluded or at least be controlled over such
relatively short look-back times? 

\section{SNe~Ia Model Considerations} \label{ModelConsiderations}
Delayed detonation models \citep{hoflich2000} predict that $m$ evolution 
may arise from (1) lower metallicities $Z$, (2) higher main-sequence masses 
$\mathfrak{M}_{\mathrm{MS}}$, and from (3) different accretion rates at high
redshifts. 

(1) The dominant effect of decreasing the metallicity $Z$ is a decrease  of the 
C/O ratio in an exploding WD with roughly solar abundance \citep[cf.][]{dominguez2002}. 
A reduction of the C/O ratio reduces the energetics of the explosion. For models 
with similar Ni production, this leaves $M_{\mathrm{V_{\mathrm{max}}}}$ virtually 
unchanged, but increases the rise and decline times. Every 1-day increase of the 
time scales causes an off-set in $\Delta m_{15}$\footnote{The decline rate 
$\Delta m_{15}$ is the change in magnitude during the first 15 days 
after $B$ maximum.} of about $-0.^{\mathrm{m}}1$. Therefore, if all apparent 
magnitudes of the SNe~Ia are reduced to a common value of $\Delta m_{15}$, a SNe~Ia 
with low C/O ratio will incorrectly be made fainter 
\citep[see][eq. 23; the $Z$-dependent $\Delta m_{15}$ corrections is only indicated if the explosion kinematics remain unchanged]{reindl2005}, 
but at the same time its line blanketing will make it bluer, causing an overestimate 
of the absorption, and hence it will  be made brighter. Thus the two effects cancel 
in first approximation. 

This conclusion is empirically confirmed by \citet{ivanov2000} who showed for
56 nearby SNe~Ia that their luminosities corrected for $\Delta\mathrm{m}_{15}$
and absorption do not significantly correlate with the distance from the
center of their host galaxies, whereby this distance is a good metallicity indicator. 
A decrease of the metallicity decreases also the $U$ and $UV$ flux by $0.^{\mathrm{m}}2$
up to $0.^{\mathrm{m}}5$ \citep{hoflich2003}. 

(2) An increase of the main-sequence mass $\mathfrak{M}_{\mathrm{MS}}$ will have 
the same effect on the O/C ratio as a decrease of $Z$, but at the same time it 
will cause lower expansion velocities, as can be measured, for instance, in the 
Si lines. A decrease of the expansion velocity by $2000\,\mathrm{km\,s}^{-1}$ will 
decrease the maximum luminosity by  $\sim0.^{\mathrm{m}}1$, while the tail 
luminosity remains unchanged \citep{dominguez2001,hoflich2003}. The distribution of 
$\mathfrak{M}_{\mathrm{MS}}$ may vary even in nearby galaxies depending on the
epoch of star formation. This effect, however, can be minimized by using only SNe~Ia 
in E/S0 galaxies (see below), whose main star formation lies generally in the distant past. 

(3) As a separate effect, the typical accretion rate may change with the epoch and, 
consequently, the central density at the time of explosion. The final kinetic 
energy of the envelope will change because the potential energy of the WD is
increasing with the central density. $M_{\mathrm{V_{\mathrm{max}}}}$ will remain
unchanged for the same amount of $^{56}\mathrm{Ni}$ produced, but the light curve 
steepens during the rise and decline phase similar as under (1) with the corresponding
effect on $\Delta\mathrm{m}_{15}$. 

Some SNe~Ia may be the product of merges, and their contribution may change
with redshift. However, this does not pose a problem because spectroscopic 
peculiarities are expected in this case \citep{hoflich1996}. Also unusual 
SNe~Ia like 1986\,G, 1991\,T, 1991\,gb, and 1999\,aa can be easily weeded out by 
their (near-maximum) spectra. 

Recent comparisons of the spectra of nearby SNe~Ia and those with redshifts of
up to $z=0.8$ do not show systematic differences \citep{matheson2005,hook2005}, 
but it is uncertain whether the expected differences would
show up in these first-level comparisons.
Changes of the initial metallicity lead also to different yields of some trace 
elements. For instance $^{54}\mathrm{Fe}$ is depressed by $\sim10\%$ in case of 
low-metallicity progenitors \citep{hoflich1998,hoflich2000}. 
\citet{bongard2005} have recently defined intensity ratios of certain spectral 
features which correlate directly with the (relative) luminosity of a SN~Ia. Si-related 
features seem particularly successful giving a scatter of only $\sim0.^{\mathrm{m}}07$.

As stated above, the decisive point for the proposed experiment is that the mean
luminosity of SNe~Ia does not change systematically between $z=0$ and $z=0.16$. 
The above discussion of the known reasons, which may cause a luminosity variation, has shown
that they cause simultaneously variations of the light curve parameters and spectra 
of SNe~Ia. If the SNe~Ia used for the experiment, or a statistically significant subset, 
show no such $z$-dependent variation of their light curve parameters and spectra it may
safely be assumed that also their mean luminosity is constant over the $z$-interval
considered, or at least that any luminosity variation is small in comparison to the 
signal of $0.^{\mathrm{m}}15$ which we are seeking. 

\section{The Experiment} \label{TheExperiment}
The possibility to detect a signal of $\Lambda$ from nearby 
($z\leq0.16$) SNe~Ia is investigated. The available high-quality 
SNe~Ia with  $z\leq0.12$ are used as a Training Set in Sect. \ref{TrainingSet}. 
Then, on the assumption that an equivalent sample of 200 SNe~Ia with 
${z\leq0.16}$ will become available, the resulting confidence 
range of $\Lambda$ is predicted in Sect. \ref{MonteCarlo}; an optimization
is discussed in Sect. \ref{Improvement}. The feasibility of the experiment is 
discussed in Sect. \ref{Feasibility}. The conclusions are given in Sect. \ref{Conclusions}.

\subsection{The Training Set} \label{TrainingSet}
A complete sample of 35 blue SNe~Ia with $1200\leq z\leq30\,000\,\mathrm{km\,s}^{-1}$
and $(B-V)<0.20$ (after correction for galactic absorption) has been
compiled by \citet{parodi2000}. Two SNe~Ia with non-normal spectra and
seven SNe~Ia which appeared in the inner parts of their spiral parent
galaxies and which show signs of internal absorption are omitted here. If, in
addition, the nine SNe~Ia are excluded which have large galactic absorption
corrections ($A_{\mathrm{V}}<0.20$) according to \citet{schlegel1998} 
the remaining 26 SNe~Ia define, on the assumption of $\Omega_{\mathrm{M}}=0.3$,
$\Omega_{\mathrm{\Lambda}}=0.7$, a Hubble diagram with a scatter of 
only $\sigma_{\mathrm{m}}=0.^{\mathrm{m}}12$ in $B$,$V$ and $I$  
after they are homogenized as to decline rate $\Delta m_{\mathrm{15}}$ and 
intrinsic color \citep{parodi2000}.

If instead the 26 SNe~Ia are fitted to a Hubble line corresponding to $\Omega
_{\mathrm{M}}=1$, $\Omega _{\mathrm{\Lambda}}=0$ the scatter increases 
marginally to $\sigma_{\mathrm{m}}=0.^{\mathrm{m}}13$. This 
suggests that the SNe~Ia contain at least some information favoring a
positive value of $\Omega_{\mathrm{\Lambda}}$. A $\chi^{\mathrm{2}}$-test 
excludes the assumption that the $\Omega_{\mathrm{M}}=1$ universe is an equally 
good fit to the data as the $\Omega_{\mathrm{M}}=0.3$, $\Omega_{\mathrm{\Lambda}}
=0.7$ universe at the $57\%$  level. The result is unsatisfactory, but implies 
that a lager sample of SNe~Ia must give a better result. The proviso is of course 
that the luminosity dispersion of homogenized normal SNe~Ia is purely statistical. 

\subsection{Monte Carlo Calculation} \label{MonteCarlo}
A random sample of 200 SNe~Ia is created with constant density in redshift
space adopting a universe with $\Omega_{\mathrm{M}}=0.3$,
$\Omega_{\mathrm{\Lambda}}=0.7$ \citep{carroll1992}. The upper limit of their redshifts 
is taken to be $z=0.16$. Their individual redshifts are used to calculate their (relative) 
apparent magnitudes under the assumption of a constant absolute magnitude 
$M_{\mathrm{\max}}$\footnote{In the present case $M_{\mathrm{\max}}^{\mathrm{B}}=-19.5$ 
and H$_{\mathrm{0}}=60\,\mathrm{km\,s}^{-1}\,\mathrm{Mpc}^{-1}$ was assumed only for illustration; 
the choice is inconsequential for what follows.}. The resulting values of $m$ were 
then degraded by a random luminosity scatter of $\sigma_{\mathrm{m}}=0$.
$^{\mathrm{m}}12$ as observed by \citet{parodi2000}. 

\begin{figure}[ht!] % positioniere hier
\centering
\includegraphics[width=9cm]{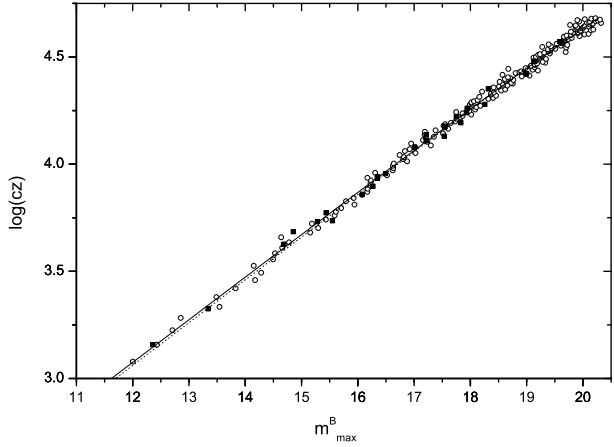}
\caption[]{The Hubble diagram of 200 randomly generated SNe~Ia with 
           $z\le0.16$ (circles) assuming a random luminosity scatter of 
           ${\sigma_{\mathrm{m}}=0.^{\mathrm{m}}12}$ (after correction for  
           galactic absorption, decline rate, and intrinsic color). The dashed line 
           is for $\Omega_{\mathrm{M}}=1$, the full line for
           $\Omega _{\mathrm{M}}=0.3$ and $\Omega _{\Lambda}=0.7$. The sample of the 
           35 observed SNe~Ia (Training Set) are represented by squares.
          }
\label{Hdiag}
\end{figure}

The generated values of $\log cz$ and $m_{\mathrm{\max}}^{\mathrm{B}}\pm\sigma$ 
are plotted in a Hubble diagram (Fig. \ref{Hdiag}). We ask now how well the
data from the "observed" 200 SNe~Ia can distinguish between a flat matter-only 
universe and a $\Omega_{\mathrm{M}}=0.3$, $\Omega_{\mathrm{\Lambda}}=0.7$ model. 
A $\chi^{\mathrm{2}}$-test shows that the latter is favored with a $89\%$ probability.

\begin{figure}[ht!]
\centering
\includegraphics[width=9cm]{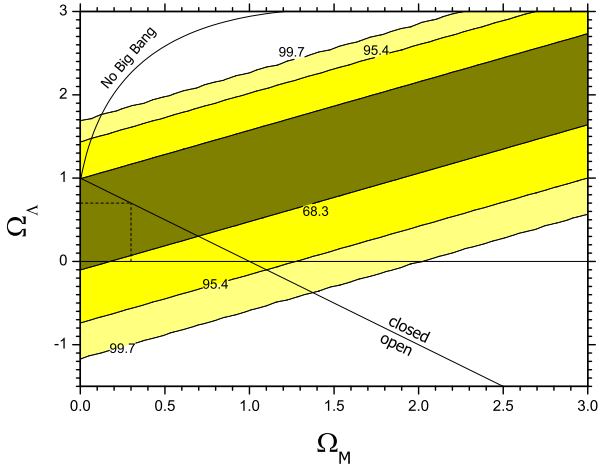}
\caption[]{Confidence contours in the $\Omega_{\mathrm{M}}$, $\Omega_{\mathrm{\Lambda}}$ 
           plane as defined by 200 randomly generated nearby SNe~Ia ($z<0.16$)
           with a luminosity scatter of $\sigma_{\mathrm{M}}=0.^{\mathrm{m}}12$.
          }
\label{200s012}
\end{figure}

A more general solution is obtained if the $1\sigma -$, $2\sigma -$ 
and $3\sigma -$confidence intervals, as defined by the $200$ randomly generated
SNe~Ia, are plotted in an $\Omega_{\mathrm{M}}$ versus $\Omega_{\Lambda}$ 
diagram (Fig.\,\ref{200s012}). It can be seen that the probability of $\Lambda$ 
being positive is lager than $68\%$ ($\approx1\sigma$) for any value of $\Omega_{\mathrm{M}}$, i.e., 
independent of the assumption of flatness. If flatness is assumed the significance is increased to $89\%$. 

\subsection{Improvement of the Experiment} \label{Improvement}
The very uneven distribution of the SNe~Ia along the Hubble line in Fig. \ref{Hdiag}
is highly unfavorable for the determination of the curvature of that line and
hence of $\Lambda$. The less than optimal distribution of the SNe~Ia is the result of 
the adopted constant space density. If it will be possible to obtain 50 SNe~Ia with
$4000\lesssim v\lesssim 7000\,\mathrm{km\,s}^{-1}$ out of a total of 200 with 
$z\leq 0.16$ (with a luminosity scatter of $\sigma_{\mathrm{M}}=0.^{\mathrm{m}}12$ as before), 
a distinction between a flat model with $\Lambda=0$ and $\Lambda=0.7$ can be 
achieved with a confidence of $95.4\%$ ($2\sigma$), instead of $89\%$ from a sample 
reflecting constant space density. Of course the discovery of these relatively nearby 
SNe~Ia is time-consuming because of the lower surface density of the nearer parent 
galaxies. SNe~Ia with $v\lesssim 3000\,\mathrm{km\,s}^{-1}$ are not useful for the experiment, 
because their expected peculiar motions of $\sim 300\,\mathrm{km\,s}^{-1}$ will substantially 
add to the statistical noise.

The experiment would be further improved if sufficient SNe~Ia are available to restrict
the analysis to events in E/S0 galaxies. Their in first approximation homologous average 
star formation, taking place predominantly at early epochs, will cause even less variations 
of the C/O ratios and metallicities of the SN\,Ia progenitors during the last $15\%$ of the 
age of the universe. Moreover, a considerable part of the observed scatter of the Hubble 
diagram must be due to unaccounted dust in late-type host galaxies. In fact 20 SNe~Ia in
E/S0 galaxies with $5000<v_{\rm{CMB}}<20\,000\,\mathrm{km\,s}^{-1}$ exhibit a scatter of only
$\sigma_{\rm{m}}=0.^{\rm{m}}10$ in $I$ \citep{reindl2005}. This seemingly modest reduction
of the scatter would considerably strengthen the detection of $\Lambda$.

\section{Feasibility} \label{Feasibility}

\subsection{Luminosity Dispersion of SNe~Ia}
The first question is whether the adopted luminosity dispersion of $0.^{\mathrm{m}}12$ is 
realistic, once the SNe~Ia are reduced to a common decline rate and color. \citet{parodi2000}
have found $\sigma_{\mathrm{B}}=0.^{\mathrm{m}}12$, but their sample is confined to
blue objects. An unrestricted sample of 62 SNe~Ia with $3000\leq v_{\mathrm{CMB}}\leq 20\,000\,\mathrm{km\,s}^{-1}$
gives a somewhat larger scatter of $\sigma_{\mathrm{B}}=0.^{\mathrm{m}}14$ or $0.^{\mathrm{m}}12$, 
much of which, however, must be due to the attempted corrections for internal absorption 
\citep{reindl2005,wang2006}. As mentioned before, a subset of 20 SNe~Ia in E/S0 
galaxies yields a scatter of only $\sigma_{\mathrm{I}}=0.^{\mathrm{m}}10$ due to the generally 
small dust content of the parent galaxies. Even this small scatter is probably still inflated by 
observational magnitude errors. The dispersion of $0.^{\mathrm{m}}12$ adopted for the present 
experiment is therefore rather conservative.

\subsection{The Frequency of SNe~Ia}
The annual frequency of detectable SNe~Ia is approximated by 
\begin{equation}
\label{eq1}
\log N=2.44+0.6\,(m_{\mathrm{limit}}-17.0) 
\end{equation}
\citep{hog1999}, where $m_{\mathrm{limit}}$ is measured in $B$ or $V$. The apparent 
maximum magnitude in $B$ or $V$ of a SN\,Ia at redshift $z$ is given by 
\begin{equation}
\label{eq2}
m_{\mathrm{max}}=5\log cz-3.465 
\end{equation}
\citep{reindl2005}. Hence an unreddened SN\,Ia at redshift ${z=0.16}$, i.e., the 
limit chosen here, will have $m_{\mathrm{B}}\approx m_{\mathrm{V}}=20.^{\mathrm{m}}0$. 
Inserting this value into (\ref{eq1}) will yield 17000 SNe~Ia per year. The number is high
enough to choose the most favorable objects in early-type galaxies ($\sim 25\%$), at high
galactic latitudes, and with the possibility to follow them up for at least 2-3 weeks post
maximum.

Here is not the place to develop a detailed strategy for the discovery and follow-up of
the SNe~Ia. It is expected that the "Nearby Supernova Factory" \citep{woodvasey2004} will
contribute many useful SNe~Ia with $0.03<z<0.08$. Also the GAIA satellite will efficiently
find about half of all SNe~Ia with $m_{\mathrm{max}}=18.^{\mathrm{m}}0$ ($z\lesssim 0.07$), 
i.e., $\sim4500$ SNe~Ia during its projected four-year lifetime \citep{tammann2002}. 
GAIA as well as Pan-STARRS \citep{kaiser2005} and Sky Mapper \citep{schmidt2005} will also be 
important for detecting the desirable, but rare SNe~Ia with 
$0.01<z\lesssim 0.02$ due to their full-sky coverage. The SNe~Ia with 
$0.07\lesssim z<0.16$ have high surface densities and will be found by ongoing
(e.g., ESSENCE, \citet{miknaitis2004}; SDSS, \citet{frieman2004}) or future searches 
for intermediate- or high-$z$ SNe~Ia. 

All SNe~Ia require excellent photometric follow-up from the ground at least in $B$, $V$, and $I$. 
Spectroscopy is desirable for as many SNe~Ia as possible. Only the spectroscopic subtypes like 
SN1991\,T and SN1999\,aa, which make up $\lesssim 15\%$ of all SNe~Ia, compete with or exceed the 
high luminosity of normal SNe~Ia \citep{reindl2005}. They have all small decline rates of 
$\Delta m_{15}<1.0$ and could be eliminated on this basis (this, however, would also
eliminate about one third of all SNe~Ia), or simply by $2\sigma$-clipping. All SNe less luminous than
normal SNe~Ia are either much fainter or - in the case of SN1989\,G - very red.

\section{Conclusions} \label{Conclusions}
The main purpose of the proposed experiment is to reduce the
possible effect of cosmic luminosity evolution of SNe~Ia on the
determination of $\Lambda$. 

If $\Lambda$ is determined from a comparison of local SNe~Ia with those at $z\approx0.8$ 
it is expected that the latter come on average from more massive and metal-poor
progenitors. If the comparison is restricted to SNe~Ia with $0<z<0.16$, as proposed here, 
it is possible to obtain their high-quality spectra and light curve parameters. If these
turn out to be independent of distance, it is a strong
argument that also their mean luminosities are (nearly) the same. 

The main conclusion is that 200 local SNe~Ia with $z\leq0.16$ can decide about  
$\Lambda>0$ for any value of $\Omega_{\mathrm{M}}$ with a significance of $68\%$, 
provided the luminosity distribution of normal SNe~Ia is Gaussian. The significance is 
increased to $89\%$ if a flat universe is assumed. The experiment can
gain considerable significance if the adopted luminosity dispersion of 
$0.^{\mathrm{m}}12$ can be reduced by the restriction to SNe~Ia in E/S0 galaxies with little
internal absorption and/or by precision photometry; it will also be possible in the future to
increase the sample size. 

It is conceivable that local and not so local SNe~Ia will eventually contribute to map the 
acceleration as a function of $z$ and thus provide a clue to the true nature of the Dark 
Energy in the universe. If, however, it will be found that the light curve parameters and/or 
spectra change systematically even within $z<0.16$, the role of SNe~Ia as cosmological 
standard candles must be rediscussed.

\smallskip
\smallskip
\smallskip
\emph{Acknowledgements.} 

The authors are grateful to Peter H\"oflich and Richard Ellis for their helpful and insightful 
comments. PE acknowledges financial support by the German \emph{Deutsche Forschungsgemeinschaft}, DFG, 
through Emmy-Noether grant Ri 1124/3-1.

\setlength{\bibsep}{3pt}   % gap between the references
\bibliographystyle{aa}
\bibliography{/aibn146_1/perni/work/latex/bibliography}

\end{document}